# Aortic pulse wave velocity measurement via heart sounds and impedance plethysmography

Roman Kusche[1][0000-0003-2925-7638], Arthur-Vincent Lindenberg[1], Sebastian Hauschild[1], and Martin Ryschka[1]

[1] Laboratory of Medical Electronics, Luebeck University of Applied Sciences, 23562 Luebeck, Germany

`roman.kusche@fh-luebeck.de; martin.ryschka@fh-luebeck.de`

**Abstract.** The determination of the physical characteristic of the human arterial system, especially the stiffness of the aorta, is of major interest for estimating the risk of cardiovascular diseases. The most common measurement technique to get information about the state of the arterial system is the pulse wave analysis. It includes the measurement of the pulse wave velocity inside the arteries as well as its morphologically changes when propagating through the arteries. Since it is difficult to detect the pulse wave directly at the aorta, most available devices acquire the pulse wave at the extremities instead. Afterwards, complex models and algorithms are often utilized to estimate the original behavior of the pulse wave inside the aorta.

This work presents an impedance plethysmography based technique to determine the aortic pulse wave velocity. By measuring the starting time of the pulse wave directly at its origin by the acquisition of heart sounds and the arrival time at the end of the aorta non-invasively via skin electrodes, unreliable complex models or algorithms aren't necessary anymore to determine the pulse wave velocity.

After describing the measurement setup and the problem-specific hardware system, first measurements from a human subject are analyzed and discussed.

## 1    Introduction

Arterial pulse wave analysis has become a useful technique in biomedical engineering in the past several years [1]. It comprises different methods like the determination of the pulse wave velocity within the arteries or the analysis of the pulse wave morphology [2]. The overall goal of all methods is to obtain information about the physical behavior of the aorta, preferable non-invasively [1].

Therefore, several measurement setups and devices have been developed in the past. Since it is difficult to measure the pressure pulse wave directly at two positions of the aorta to obtain its velocity, more practical measurement methods have been developed. Nowadays, a common method is the acquisition of the pulse wave at only one single point of the body which is easily accessible, like the radial artery or the brachial artery [3]. The morphology of the pulse wave within this vessel is then acquired by the usage of pressure sensors in various specific measurement setups, e.g. by applying a blood pressure cuff to the upper arm. The actual interesting information, the pulse wave velocity within the aorta, is then estimated by complex algorithms and models [4].

One problem of this procedure is to find a reliable model since the physical behavior of the aorta and of other arteries are very different [5]. Additionally, this model has to fit to the individual subject's anatomy. Furthermore, the measurement of the pulse wave by applying a pressure to the arteries, as performed by cuff-based systems, influences the behavior of the measured vessel itself.

The approach of this work is to measure directly the arterial pulse wave velocity within the aorta. Therefore, the starting time of the pulse wave is determined via the acquisition of the heart sounds. The arrival time is detected by performing a bioimpedance measurement at the lower leg.

In this work the acquired biosignals are introduced and the measurement approach is explained in detail. Afterwards, the developed measurement system is described with a focus on the signal processing. Finally, first measurements from a human subject are presented and discussed.

## 2 Materials and Methods

### 2.1 Phonocardiography

Phonocardiography is the measurement of the heart sounds from a subject. Therefore, a microphone or a stethoscope can be used. The most conspicuous characteristics of this signal are the $1^{st}$ and the $2^{nd}$ heart sound. These sounds occur, when the mitral valve, respectively the aortic valve, is closing. Although the $1^{st}$ heart sound does not exactly represent the opening of the aortic valve, it occurs directly before the actual pulse wave starting time and is therefore a useful information [6]. Furthermore, this signal is related to the actual heart muscle contraction in contrast to an ECG signal, which just represents the electrical stimulation of the heart.

### 2.2 Impedance Plethysmography

Impedance plethysmography is the acquisition of arterial pulse waves by performing electrical bioimpedance measurements. This technique is based on the fact, that the pressure pulse wave provokes also changes in the diameter of the arteries and therefore affects the amount of blood volume within the tissue under test. Since the conductivity of blood is significantly higher than that of other tissue, these minuscule volume changes can be detected by performing a continuous bioimpedance measurement [7]. Assumable impedance magnitude changes are in the range of Milliohms, when performing bioimpedance measurements at the extremities [8].

### 2.3 Measurement Approach

The overall goal of this approach is the determination of the pulse wave velocity (PWV), especially within the aorta. Therefore, the pulse wave is detected at two different points, as illustrated in Fig. 1. As starting point, the heart is chosen and the pulse wave starting time is defined by the 1st heart sound, which is detected via phonocardiography with a digital stethoscope.

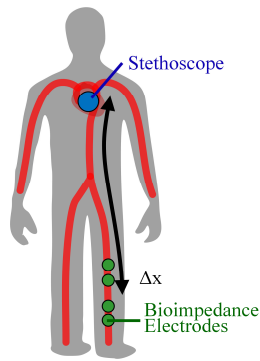

**Fig. 1.** Measurement setup to detect the starting time and the arrival time of the pulse wave for calculating the average pulse wave velocity.

The pulse wave arrival time is detected at the lower leg of the subject by performing a continuous bioimpedance measurement. Obviously, the arrived wave has passed the aorta but also a piece of the femoral artery. However, the measurement setup avoids discomfort for the subject. This compromise is accepted in this approach.

With the obtained information about the starting time ($t_{Heart}$) and the arrival time ($t_{Leg}$) of the pulse wave and the additional knowledge about the covered distance $\Delta x$, the average pulse wave velocity (PWV) can be calculated with Formula 1.

$$PWV = \frac{\Delta x}{t_{Leg} - t_{Heart}} \tag{1}$$

### 2.4 System Development

To implement the proposed measurement approach, a measurement system has been developed. As shown in Fig. 2, this system consists of an analog part (yellow), which includes sensors and electronic circuits, and a digital (green) part, implemented in Matlab (from Mathworks). The upper signal path represents the extraction and detection of the 1st heart sound to determine the pulse wave starting time. The lower path is intended to determine the pulse wave arrival time at the lower leg.

The heart sounds are acquired via a digital stethoscope, realized by an electret microphone capsule. To remove the undesired DC offset in the microphone signal, the heart sounds are high pass filtered. Afterwards, the signal is amplified and digitized with a sampling rate of 1000 SPS. To determine the maxima of the heart sounds, the signal is enveloped on the digital signal processing side and a peak detection is performed. The corresponding points in time, which represent the 1st heart sounds are stored as pulse wave starting times.

To determine the pulse arrival times at the leg, the bioimpedance is measured via surface electrodes. A small known AC current is applied via a voltage controlled current source (V/I) into the tissue under observation ($Z_{Bio}$). The occurring voltage drop is amplified and afterwards an amplitude modulation (AM) demodulator extracts the actual information about the impedance magnitude. This information is digitized via a second, synchronized ADC channel and for attenuating high frequency noise low pass filtered. In order to extract the pulse arrival time, the inflection point before the actual pulse wave is determined by differentiating the signal twice [9]. This point indicates the beginning of the pulse wave morphology change and therefore the beginning of the arriving pulse wave.

Finally, the obtained pulse wave starting times are subtracted from the corresponding arrival times to get the pulse transient times.

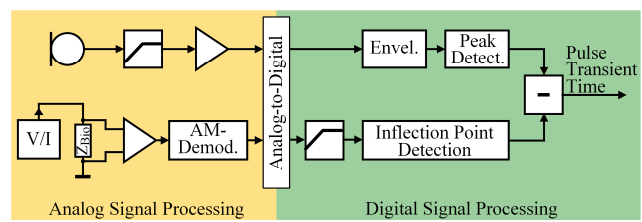

**Fig. 2.** Block diagram of the developed pulse wave velocity measurement system, consisting of an analog and a digital signal processing part.

The analog signal path was implemented as an application specific printed circuit board (PCB), as shown in Fig. 3 (a). The PCB has a size of 152 x 84 mm² and contains more than 450 components. It is capable of acquiring the heart sounds as well as two independent bioimpedance channels, using two additional current source boards, which are also depicted in the photograph. To fulfill the electrical safety requirements of the standard for medical electrical equipment (IEC 60601-1), medical power supplies, as well as a galvanically isolated USB interface were implemented.

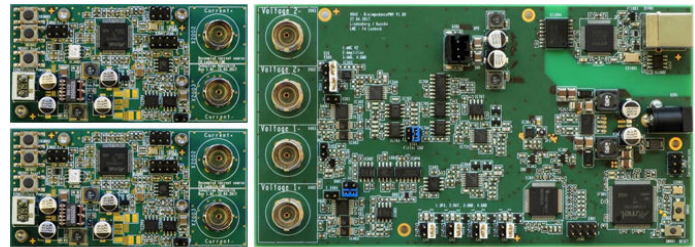

(a)

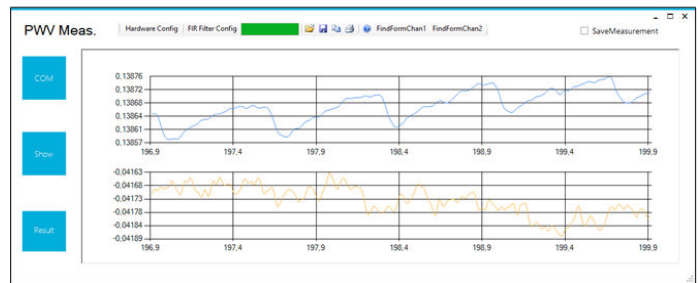

(b)

**Fig. 3.** Photograph of the pulse wave velocity measurement system and the corresponding real-time user interface.

In Fig. 3 (b) a screenshot of the developed real-time data acquisition software, programmed in C# language is shown. It can be utilized to configure the hardware system and to monitor the acquired signals during a measurement procedure. Afterwards, the data can be exported to Matlab-compatible file format.

## 3    Results

An exemplary measurement on a human subject for a duration of 8 s has been performed, using the developed hardware system as well as the proposed digital signal processing steps. In Fig. 4 the resulting envelope (blue) of the heart sound signal is plotted and the detected peaks, caused by the 1$^{st}$ heart sounds, are marked (orange).

In the plot below, the measured and filtered bioimpedance signal is presented. For better display, this signal has been inverted. It shows the typical pulse wave morphologies. The obtained pulse wave arrival times at the leg are marked (purple) as well.

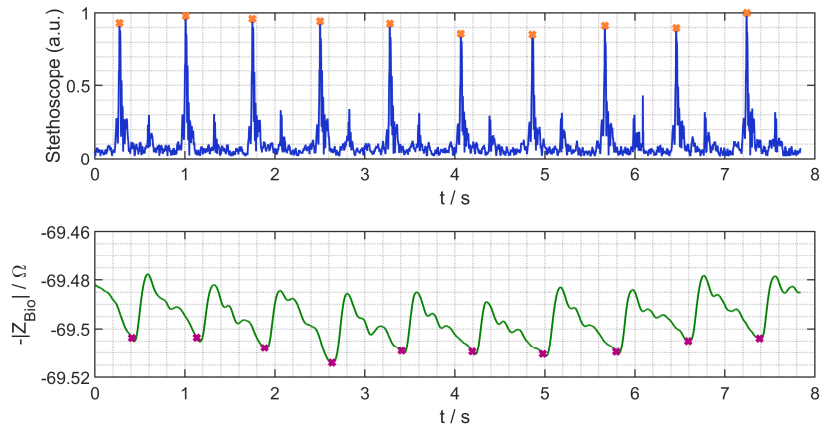

**Fig. 4.** Envelope of the acquired heart sounds (blue), and the measured bioimpedance (green). The crosses in the plots indicate the obtained points in time, in which the pulse wave starts and arrives, respectively.

Subtracting of the corresponding detected points in time from each other leads to the pulse transient times. Averaging the calculated pulse transient times of these 10 pulse waves and assuming the distance Δx to be 1.35 m leads to a mean pulse wave velocity of 10.3 m/s (SD=±0.6 m/s), using Formula 1. An additional system characterization has proven that time delays, generated by the measurement system, are lower than 5 ms in the relevant frequency range.

## 4 Discussion

A first measurement from a human subject has demonstrated the usability of the proposed measurement approach. Due to the influence of the femoral artery to the measured average pulse wave velocity, the obtained value does not only represent the velocity inside the aorta. It is assumed, that this issue can easily be solved by performing a second bioimpedance plethysmography at the same leg. Thereby, the velocity inside the femoral artery could be determined and with the knowledge about the length of the aorta, both the velocities can be separated from each other.

Another point of uncertainty is the usage of the heart sounds. As described before, this event does not physically correspond to the actual opening of the aortic valve. Implementing empirical values of this time delay to the proposed algorithm could improve the results.

A common problem of detecting the pulse wave at specific points, like the minimum, the maximum or the inflection point of the propagating wave, is that these points are based on the phase delay. However, the group delay is the actual value of interest.

## Acknowledgement

This work has been supported by the German Federal Ministry of Education and Research (BMBF) under the project LUMEN II (FKZ01EZ1140A/B).

## Conflicts of Interest

The authors declare that they have no conflict of interest.